\documentclass[twocolumn,showpacs,preprintnumbers,amsmath,amssymb,prl]{revtex4}

\usepackage{graphicx}
\usepackage{dcolumn}
\usepackage{bm}

\begin{document}

\title{Temperature-dependent screening of the edge state around antidots \\
in the quantum Hall regime}

\author{Masanori Kato}
\email{masanori@issp.u-tokyo.ac.jp}

\author{Akira Endo}%
\author{Shingo Katsumoto}%
\author{Yasuhiro Iye}%

\affiliation{Institute for Solid State Physics, University of Tokyo, 5-1-5 Kashiwanoha, Kashiwa, Chiba 277-8581, Japan}%

\date{\today}

\begin{abstract}
The Aharonov-Bohm (AB) effect in a small array of antidots with large aspect ratio is investigated in the quantum Hall regime.
The evolution with temperature of the AB oscillations in the magnetic field vs gate voltage ($B$-$V_\mathrm {g}$) plane reveals the temperature dependence of the screening. 
The self-consistently screened potential of the compressible band surrounding an antidot is observed to gain progressively steeper slope with increasing temperature. 
\end{abstract}

\maketitle

The integer quantum Hall (QH) effect, observed in two-dimensional electron gas (2DEG) subjected to a strong perpendicular magnetic field~\cite{Prange}, is generally explained within a non-interacting electron picture.
In such a picture, the edge state along the system boundary is simply envisaged as the crossing point of the Landau level (LL), which rises in energy as the boundary is approached, with the Fermi level ($E_{\mathrm F}$).
The width of the edge states is then determined by the spatial extension of the relevant single particle wave function, which is given by the magnetic length $\ell = (\hbar/eB)^{1/2}$. 
When the screening effect of a 2DEG is taken into consideration, the LLs take a more complicated profile in the vicinity of $E_{\mathrm F}$.
The energy dispersion of the edge states becomes stepwise, consisting of compressible and incompressible regions at zero temperature \cite{Chklovskii}.
Theoretical calculations extended to finite temperatures \cite{Lier, Oh} predict progressive smearing of the stepwise profile with increasing temperature.
Thus the temperature dependence of the edge potential profile reflects the self-consistent screening of 2DEG at finite temperatures. 
Experimentally, the temperature-induced change of the edge state was invoked by Machida {\it et al.}~\cite{Machida} in their interpretation of the inter-edge-channel transport. 
In order to attain more quantitative account of the evolution of the edge states with temperature, 
the present work addresses this issue via study of the Aharonov-Bohm (AB) effect in an antidot system. 

A quantum antidot, i.e., a submicron potential hill for 2D electrons, constitutes an artificial \lq\lq boundary" and provides a useful system for the study of the QH edge states. 
Under a strong perpendicular magnetic field, electrons form bound states around the antidot. 
The magnetotransport of antidot systems in the QH regime ubiquitously exhibits AB oscillations (ABOs) \cite{Ford, Kataokadf, Kataokakondo, Karakurt, Camino05, Goldman08, Katosingle}, which arise from coupling of the localized orbits to the current carrying extended states. 
Part of magnetocounductance features, e.g., oscillation period, can be understood within a single-particle picture. 
However, many experiments \cite{Ford, Kataokadf, Kataokakondo, Goldman08, Katosingle} and theories \cite{Sim, Zozou} point to the crucial role played by the electron-electron interaction. 
A full understanding of the antidot systems therefore requires detailed knowledge of the structure of edge states which inevitably involves the self-consistent screening effect.
Conversely, detailed experiments should furnish relevant information on the QH edge states.

In a preceding paper~\cite{RFAB}, we have shown that the evolution of the ABOs as a function of magnetic field and carrier density reveals the strong modification of the potential profile around an antidot by the screening effects. 
Here we focus on the temperature dependence of the ABOs in the magnetic field-density plane. 
We show that the slope of the screened potential around an antidot becomes steeper with increasing temperature. 
This can be attributed to the temperature dependence of the screening effect as predicted by the existing theoretical calculations~\cite {Lier, Oh}.

Antidot array samples were fabricated from a GaAs/AlGaAs single heterojunction wafer (density $n=4.0\times 10^{11}~\mathrm{cm^{-2}}$ and mobility $\mu =98~\mathrm{m^2/Vs}$). 
The sample used in this study was a $5.3~\mathrm{\mu m}$ wide Hall bar containing $5\times 5$ antidots in a square lattice pattern used in the previous work~\cite {nu2, RFAB}.
The lattice period was $a=1~\mathrm{\mu m}$ and the antidot radius was $r =350~\mathrm{nm}$. 
The effective radius $r^*$ of the antidot was larger than the lithographical radius $r$ typically by $\sim 100~\mathrm{nm}$, so that the channels between two neighboring antidots were $\sim 100~\mathrm{nm}$ wide at their narrowest point.
This sort of antidot array system can be viewed as a network of a narrow conducting channels. 
The ohmic contacts to the Hall bar were made with AuGe/Ni electrode pads.
A Au-Ti Schottky front gate enabled us to tune the Fermi energy of the system. The resistance measurements were made by a standard low-frequency ($13~\mathrm{Hz}$) lock-in technique with low excitation current (typically $1~\mathrm{nA}$) at different temperatures ranging from 30 to 800~mK. 

\begin{figure}[htb]
\begin{center}
\includegraphics[width=0.95\linewidth]{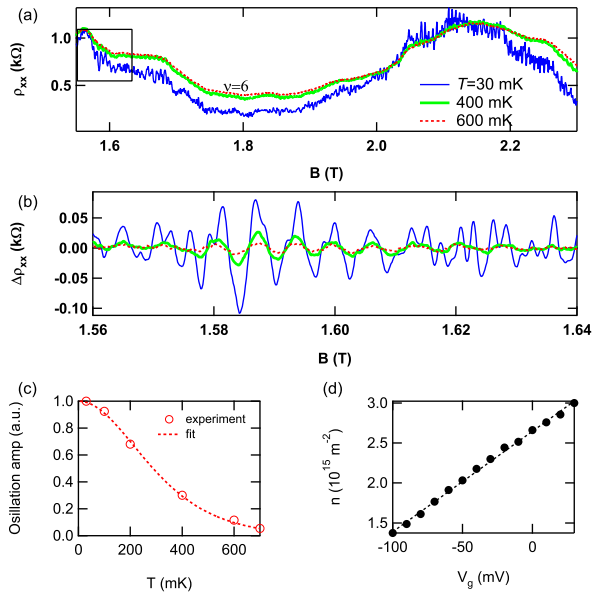}
\caption{(Color online) (a) $\rho _{xx}$ at $V_\mathrm {g} = 0~\mathrm{mV}$ around $\nu=6$ for different temperatures. 
(b) The oscillatory component of the trace in the square area in (a) obtained by subtraction of the smooth background. 
(c) Temperature dependence of the amplitude of the periodic oscillations around $B=1.6$ T. 
Open circles represent the experimental data. Dashed line is the fit to the Dingle function $aT/\sinh (aT)$ ($a=2\pi ^2k_\mathrm {B}/\Delta E,~\Delta E=0.22~\mathrm {meV}$). 
(d) The dependence of the density on the gate voltage $V_\mathrm {g}$ in the antidot sample obtained from the magnetic field positions of the center of the Hall plateaus as a function of $V_\mathrm {g}$. 
The dependence is linear in the range of $V_\mathrm {g}$ studied; the conversion factor $\Delta n/\Delta V_\mathrm {g}=0.013\times 10^{15}~\mathrm {m^{-2}/mV}$ is obtained.}
\label{FIG1}
\end{center}
\end{figure}

Figure~\ref{FIG1}(a) shows the magnetoresistivity $\rho _{xx}$ of the sample at $V_\mathrm {g}=0$ mV around the LL filling factor $\nu =6$ ($B\simeq 1.8$ T) for different temperatures.
The trace contains an oscillatory component $\Delta \rho _{xx}$ in the QH transition region as shown in Fig.~\ref{FIG1}(b). 
The period of the oscillation is $\Delta B\simeq 5.7~\mathrm {mT}$ corresponding to the AB period expected for the effective antidot area $S=\pi r^{*2}$ with $r^*=480~\mathrm {nm}$. 
Figure \ref{FIG1}(c) shows the decrease of the oscillation amplitude with temperature following the Dingle function, in accordance with our previous observation~\cite{Katophysica} .

\begin{figure}[htb]
\begin{center}
\includegraphics[width=0.95\linewidth]{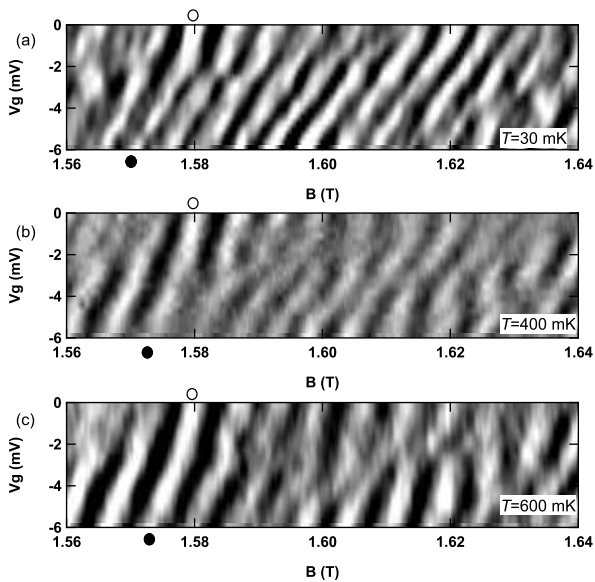}
\caption{[(a)-(c)] Gray-scale plot of $\Delta \rho _{xx}$ obtained by magnetic field sweeps with various values of $V_\mathrm{g}$. White regions represent higher resistance. 
Open and solid circles in each figure show the magnetic field position of the peak at the $V_\mathrm{g}=0$ and $-6$ mV, respectively.}
\label{FIG2}
\end{center}
\end{figure}

\begin{figure}[htb]
\begin{center}
\includegraphics[width=0.95\linewidth]{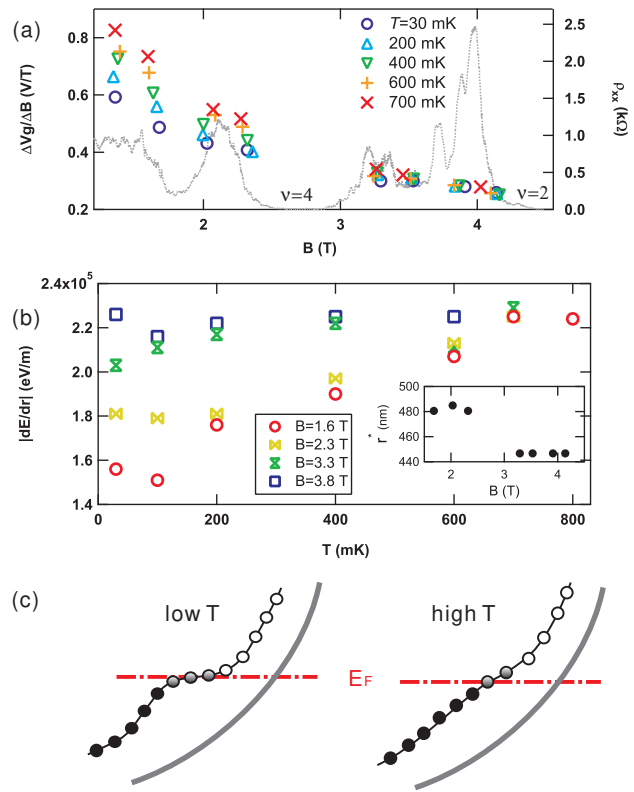}
\caption{(Color online) (a) The slopes of AB oscillations $\Delta V_\mathrm{g}/\Delta B$ at $V_\mathrm{g}=0$ mV for different temperatures. 
Each symbol is obtained after sufficient averaging over the field range $\sim 0.2$ T. Dotted gray line indicates $\rho _{xx}$ as a guide for eye. 
(b) The temperature dependence of the potential slope around an antidot for different magnetic fields obtained from the data in  (a) and $r_m(=r^*)$ at each magnetic field using Eq.~(\ref {eq:dB}). 
The inset is the effective antidot radius $r^*$ deduced from the period of the ABOs at each magnetic field. 
(c) Schematic representations of the potential near $E_{\mathrm F}$ around an antidot with the screening effects at low and high temperatures. 
The circles represent the single-particle states. The filled, half-filled and open circles correspond to the occupied, partially occupied and unoccupied states, respectively.}
\label{FIG3}
\end{center}
\end{figure}

Figure~\ref{FIG2} shows the evolution of the ABOs in Fig.~\ref{FIG1}(b) with the gate bias. 
The light and dark regions represent local peaks and dips of $\Delta \rho _{xx}$ respectively. 
With increasing temperature, the trajectories of the ABOs in the $B$-$V_\mathrm {g}$ plane becomes steeper, i.e., the slope $\Delta V_\mathrm {g}/\Delta B$ of the stripes becomes larger. 
For example, the peak marked by an open circle at $V_\mathrm {g}=0$ mV stays at the same magnetic field position $B=1.58$ T for different temperatures, while the corresponding 
peak marked by a solid circle at $V_\mathrm {g}=-6~\mathrm {mV}$ is seen to shift to higher magnetic field with increasing temperature.
As discussed in our previous paper~\cite {RFAB}, the value of $\Delta V_\mathrm {g}/\Delta B$ reflects the the screened potential profile around the antidot at $E_{\mathrm F}$. 
Let us derive the relevant relation from the Bohr-Sommerfeld quantization condition $\pi r_m^2B=mh/e$ which determines the period of the ABOs.
Here, the integer $m$ specifies the $m$-th single particle state encircling the antidot, and $r_m$ is its radius. The ratio $\Delta V_\mathrm {g}/\Delta B$ is given by

\begin{eqnarray}
\frac {\Delta  V_\mathrm {g}}{\Delta B} &=& - \frac {r_m}{2 B} \left(\frac {\Delta r_m}{\Delta V_\mathrm {g}}\right)^{-1} \cr
&=& - \frac{r_m}{2 B} \left( \frac{\mathrm{d} E}{\mathrm{d} r} \right)_{r_m} \left(\frac{\Delta E_\mathrm{F}}{\Delta V_\mathrm {g}}\right)^{-1},
\label{eq:dB}
\end{eqnarray}
where $\Delta r_m/\Delta V_\mathrm {g}$ denotes the change by the gate bias of the radius $r_m$ at $E_\mathrm {F}$ enclosing $m$ magnetic flux quanta. 
Equation (1) allows us to translate the measured slope $\Delta V_\mathrm {g}/\Delta B$ into $|dE/dr| _{r_m}$ at $E_\mathrm {F}$ by using the value $\Delta E_\mathrm {F}/\Delta V_\mathrm{g}=0.046~\mathrm {eV/V}$ 
estimated from Fig.~\ref{FIG1}(d). 
Thus the temperature dependence of $\Delta V_\mathrm{g}/\Delta B$ reveals the evolution of the potential profile around an antidot with the temperature. 
Figure~\ref{FIG3}(a) presents the $\Delta V_\mathrm{g}/\Delta B$ for different temperatures. 
As can be seen in Fig.~\ref {FIG2}, the slope slightly fluctuates from line to line, disturbed by fluctuations in the background resistance unrelated to the ABOs; 
the disturbance is more pronounced for higher temperatures where the amplitudes of the ABOs are smaller. 
To avoid the trend that takes place in a larger magnetic-field scale from being obscured by this fluctuation, we plot in Fig.~3(a) the average of the slopes from more than 20 lines 
encompassed by the magnetic-field range $\pm 0.1~\mathrm {T}$ (roughly corresponding to the size of the symbols) of the point plotted in the figure. 
The value of  $\Delta V_\mathrm{g}/\Delta B$ is observed to increase with temperature, with the increment more noticeable for lower magnetic fields and almost absent for $B\geq 4~\mathrm {T}$. 
Qualitatively the same behavior is observed around other gate biases. 

Figure~\ref {FIG3}(b) shows the temperature dependence of the potential slope $|dE/dr| _{r^*}$ at $E_F$ obtained from the values of $\Delta V_\mathrm{g}/\Delta B$ in Fig.~\ref{FIG3}(a) with $r_m(=r^*)$ calculated from the period of the ABOs. 
In all cases, the potential slope $| \mathrm{d}E/\mathrm{d}r| _{r^*}$ increases with increasing temperature.
However the effect of temperature is stronger at lower magnetic fields (higher fillings).

Let us take, as an example, the field range around $B\simeq 1.6~\mathrm {T}$.
Here the value of $| \mathrm{d}E/\mathrm{d}r| _{r^*}$ changes by 40 $\%$ from $1.6~\times10^5~\mathrm{eV/m}$ at 30 mK to $2.2~\times10^5~\mathrm{eV/m}$ at 800 mK.
These values can be compared with a different line of estimation.
The radial separation $\Delta r_m$ between two adjacent single-particle states, as estimated from $\Delta r_m \simeq  h/2\pi  er_mB$, is $0.9~\mathrm {nm}$ at $B=1.6~\mathrm{T}$. 
The energy separation $\Delta E$ between the single-particle states, on the other hand, has been estimated from the temperature dependence of the oscillation amplitudes in  Fig.~\ref{FIG1}(c) as $0.22~\mathrm{meV}$.
Combination of these values gives $| \mathrm{d}E/\mathrm{d}r| _{r^*} \simeq 2~\times 10^5~\mathrm{eV/m}$ roughly in agreement with the above values obtained. 

The change of $| \mathrm{d}E/\mathrm{d}r| _{r^*}$ with temperature is attributed to the temperature-dependent screening of the electrostatic potential for the edge states around the antidot as  exaggeratedly depicted in Fig.~\ref{FIG3}(c). 
The potential profile at the edge of the antidot is greatly modified from the bare potential by the strong screening effect \cite{Chklovskii}.
Theoretical calculations~\cite{Chklovskii, Lier} suggest that the effective potential in the compressible region becomes nearly flat at low enough temperatures.
With increasing temperature, the screening effect diminishes so that the gradient of the self-consistent potential gradually increases towards the bare value.
The theoretical calculation~\cite{Lier} indicates that substantial deviation from the low temperature limit occurs typically at $k_\mathrm{B}T/\hbar \omega_\mathrm{c} \geq 0.01$, where $\hbar \omega_\mathrm{c}$ is the LL spacing. 
The higher effectiveness of screening at higher magnetic fields is attributable to higher density of states in the relevant LL. 
The value of $k_\mathrm{B}T/\hbar \omega_\mathrm{c}$ at the highest temperature ($T \sim 800~\mathrm{mK}$) of the present experiment is $\sim 0.01$ for $B \sim 4~\mathrm{T}$.
Thus the finite-temperature reduction of the screening effect is not discernible at this magnetic field (for $T > 800~\mathrm{mK}$, the ABO is no longer visible).
At lower magnetic fields, the finite temperature effect is visible in the range of the present experiment as seen in Fig.~\ref {FIG3}(b). 

Now we discuss the value of $|\mathrm{d}E/\mathrm{d}r| _{r^*}$ qualitatively.  
Firstly, unlike the theoretical result by Lier~\cite{Lier}, the observed $|\mathrm{d}E/\mathrm{d}r| _{r^*}$ remains non-zero at $T \rightarrow 0$, i.e., the screened potential in the compressible region is not completely flat. 
This is in qualitative agreement with the calculation by Suzuki and Ando~\cite{Suzuki} of the edge states in quantum wires. 
They demonstrated that the complete flattening of the compressible stripes does not appear in a narrow wire with width $100\sim 200~\mathrm {nm}$. 
As mentioned before, our sample can be regarded as a network of the wire with the width $\sim 100~\mathrm {nm}$. 

Secondly, the value of $| \mathrm{d}E/\mathrm{d}r| _{r^*}$ in Fig.~\ref {FIG3}(b) tends to be larger at higher fields at the base temperature $T=30~\mathrm {mK}$. 
The origin of this behavior is not clear at the moment but may be related to the radius of the edge state around an antidot. 
A numerical calculation using density functional theory \cite {Zozou} showed that the compressible band of the edge states formed around an antidot tends to be narrower for a smaller antidot. 
The inset of Fig.~\ref{FIG3}(b) shows the radii of the edge states around an antidot with a sudden drop in the vicinity of the $\nu =4$ ($B\simeq 2.7~\mathrm {T}$). 
With the increase of the magnetic field, the LL that consititutes the outermost edge state around the antidot changes from the second LL to the lowest LL at $\nu =4$ accompanied by the reduction of $r^*$.  
The larger value of $| \mathrm{d}E/\mathrm{d}r| _{r^*}$ at higher magnetic fields may, therefore, be ascribable to the smaller antidot radius $r^*$. 

\begin{figure}[htb]
\begin{center}
\includegraphics[width=0.95\linewidth]{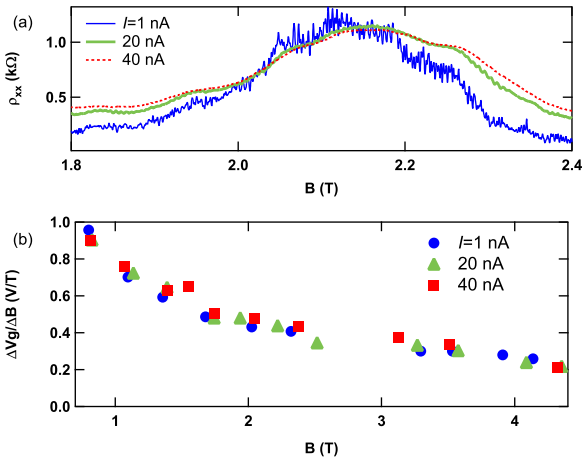}
\caption{(Color online) (a) $\rho _{xx}$ traces for $V_\mathrm{g}=0~\mathrm {mV}$ in the filling factor range between $\nu =$ 4 and 6 measured with three excitation currents at $T=30$~mK.
(b) The magnetic field dependence of $\Delta V_\mathrm{g}/\Delta B$ for different currents.}
\label{FIG4}
\end{center}
\end{figure}

Next we compare the above result of finite temperature effect with the effect of {\it electron heating} due to applied currents. 
Figure~\ref {FIG4}(a) shows $\rho_{xx}$ at $T=30$ mK for different excitation currents. 
As the current is increased, the amplitude of the ABO diminishes. 
This is similar in appearance to the case of raised temperature shown in Fig.~\ref{FIG1}(a). 
The value of $\Delta V_\mathrm{g}/\Delta B$ in Fig.~\ref{FIG4}(b), however, does not change with the bias current, implying that the effective potential around the antidot is not affected by electron heating.
The effective electron temperature can be estimated from the comparison of the amplitude of Shubnikov-de Haas oscillation or that of the ABO for each current with those at different lattice temperatures. 
The estimated values of the electron temperature are $T_\mathrm{e} \sim 400$ and $800~\mathrm{mK}$ at $I=20$ and $40~\mathrm{nA}$, respectively. 

As discussed in earlier publications~\cite{Iyeantidot, nu2}, the ABO is attributed to the coupling between the localized states around the antidot and the conductive extended states at $E_{\mathrm F}$ in the QH transition regions. 
The electron heating caused by the applied current smears out the energy distribution function of the extended states and thus the amplitude of the ABO decreases. 
However the electron heating seems not to affect the screening, in the localized states around an antidot significantly.
This can be attributed to the weakness of the coupling between the current carrying states and the edge states localized around the antidots, 
with which the current in the extended states is unable to raise the electron temperature in the localized edge states. 

In summary, we have investigated the temperature dependence of the potential profile around an antidot in the integer QH transition regions. 
The experimental results strongly suggest that the energy gradient of edge states, which is stepwise (but not flat) with the screening effects at low temperatures, smooths with increasing temperature.

\begin{acknowledgments}
This work was supported by the Grant-in-Aid for Scientific Research from the Ministry of Education, Culture, Sports, Science and 
Technology (MEXT) Japan, and was also partly supported by Special Coordination Funds for Promoting Science and Technology. 
One of the authors (M.K.) appreciates the support by the Research Fellowship for Young Scientists from the Japan Scociety for the Promotion of Science (JSPS).
\end{acknowledgments}

\end{document}